\documentclass[fleqn,usenatbib]{mnras}

\usepackage{newtxtext,newtxmath}

\usepackage[T1]{fontenc}
\usepackage{ae,aecompl}

\usepackage{graphicx}	
\usepackage{amsmath}	
\usepackage{xcolor}

\title[Impact of flares and CMEs on HD189733b]{The impact of coronal mass ejections and flares on the atmosphere of the hot Jupiter HD189733b}
\author[Hazra et al.]{
Gopal Hazra,$^{1,2}$\thanks{E-mail: hazra@strw.leidenuniv.nl}
Aline A. Vidotto,$^{1,2}$
Stephen Carolan,$^{2}$
Carolina Villarreal D'Angelo,$^{3}$
Ward Manchester$^4$
\\
$^{1}$Leiden Observatory, Leiden University, 2300 RA, Leiden, The Netherlands\\
$^{2}$School of Physics, Trinity College Dublin, Dublin 2, Ireland\\
$^{3}$Instituto de Astronom{\'i}a Te{\'o}rica y Experimental (IATE-CONICET). Laprida 854, C{\'o}rdoba, Argentina\\
$^4$Department of Climate and Space Sciences and Engineering, University of Michigan, Ann Arbor, MI48109, USA
}

\date{Accepted XXX. Received YYY; in original form ZZZ}

\pubyear{2021}
\begin{document}
\label{firstpage}
\pagerange{\pageref{firstpage}--\pageref{lastpage}}
\maketitle

\begin{abstract}
High-energy stellar irradiation can photoevaporate planetary atmospheres, which can be observed in spectroscopic transits of hydrogen lines. For the exoplanet HD189733b, multiple observations in the Ly-$\alpha$ line have shown that atmospheric evaporation is variable, going from undetected to enhanced evaporation in a $1.5$-year interval. Coincidentally or not, when HD189733b was observed to be evaporating, a stellar flare had just occurred 8h prior to the observation. This led to the question on whether this temporal variation in evaporation occurred due to the flare, an unseen associated coronal mass ejection (CME), or even the effect of both simultaneously. In this work, we investigate the impact of flares (radiation), winds and CMEs (particles) on the atmosphere of HD189733b using 3D radiation hydrodynamic simulations of atmospheric evaporation that  self-consistently include stellar photon heating. We study four cases: first- the quiescent phase of the star including stellar wind, second- a flare, third- a CME, and fourth- a flare that is followed by a CME. Compared to the quiescent case, we find that the flare alone increases the evaporation rate by only 25\%, while the CME leads to a factor of 4 increase in escape rate. We calculate Ly-$\alpha$ synthetic transits and find that the flare alone cannot explain the observed high blueshifted velocities seen in the Ly-$\alpha$ observation. The CME, however, leads to an increase in the velocity of the escaping atmosphere, enhancing the transit depth at high blueshifted velocities. While the effects of CMEs show a promising potential to explain the blueshifted line feature, our models are not able to fully explain the blueshifted transit depths, indicating that they might require additional physical mechanisms.
\end{abstract}




\section{Introduction}
Close-in exoplanets can be exposed to high levels of stellar radiation from their host stars. The high energy part of the stellar radiation (e.g., X-ray and extreme ultra-violet) drives  planetary outflow by photoionising the planetary material \citep[ e.g.,][]{2007P&SS...55.1426G, Murray-Clay2009}. As a result, the atmosphere eventually escapes from the planet, consistent with what has been observed in close-in gas planets \citep[ e.g.,][]{Vidal-Madjar2003,2010ApJ...714L.222F, Lecavelier12,Lavie2017}.  The escaping atmosphere then interacts with the  wind of the host star and depending on the strength of the stellar wind ram pressure, the atmospheric escape rate can be affected \citep[e.g.,][]{Tripathi2015, Khodachenko2015, Christie2016, Carroll-Nellenback2017, Villarreal2018, McCann19, Debrecht2020, Odert2020, Vidotto2020, Carolan2020, Carolan2021, Shaikhislamov2021}.
In the case of dense stellar wind, planetary atmospheres can be eroded by stellar wind for close-in planets \citep[e.g.,][]{Cohen2015, Rodriguez-Mozos2019}. However, recently \citet{Vidotto2020}  and \citet{Carolan2021} investigated some circumstances where stellar wind can prevent atmospheric escape.

In most of the studies carried out so far, the stellar wind has been considered as a steady, time-independent, flow interacting with the atmosphere of the planet. However, in addition to the stellar wind,  transients events (e.g., flares and coronal mass ejections, CMEs) can play a very important part in the dynamics of the atmospheric escape. These transient events change the stellar environment in two ways. On one hand, flares enhance the stellar radiation, which ionises more planetary material resulting in strong photoevaporation from the planet \citep[e.g.,][]{Chadney2017,Bisikalo2018, Hazra2020}. On the other hand, coronal mass ejections change the stellar and planetary wind conditions rapidly, which strongly affects atmospheric escape \citep[ e.g.,][]{Kay2016, Cherenkov2017}.

Most of the exoplanets discovered so far are orbiting main-sequence, low-mass  stars, from spectral types F to M. \citet{Audard2000} found that, in cool stars, the  occurrence rate of flares  with energies larger than 10$^{32}$ erg increases with X-ray stellar luminosity. Given that X-ray stellar activity  increases with  rotation rate \citep{Wright11}, fast rotators i.e., young host stars,  flare more frequently than a star of solar age \citep{Maehara2015}. This can have an important effect on the evolution of the planetary atmosphere.

Although we now have some understanding on flare occurrence, the occurrence rate of CMEs for  stars other than the Sun is currently not available. In the Sun, for example, observations of  CMEs during the solar cycle 23
show that the CME occurrence rate varies from $\sim$ 0.5 per day near solar minimum to $\sim$ 6 near solar maximum, resulting in up to more than 13000 CMEs over the cycle 23 \citep{Gopalswamy2003, Yashiro2004}. It is believed that young and more active stars could have a higher rate of CMEs \citep[e.g.,][]{Aarnio2012,Drake2013,Odert2017}. Such a frequent rate of CMEs from host stars might have an important role on planetary atmospheres and their evolution. In particular,  for close-in planets, the effects of CMEs might be more important than for planets in the solar system (at large orbital distances). This is because the CME has a higher density and magnetic energy density the closer it is to the host star \citep{Okane2021, Scolini2021}.

In this paper, we study the impact of CMEs and flares on the atmosphere of HD189733b using a self-consistent 3D radiation hydrodynamics model. Atmospheric escape in HD189733b has been  detected using transmission spectroscopy \citep{Lecavelier2010, Lecavelier12, BL2013}. In particular, \citet{Lecavelier12} observed two transits of HD189733b on two epochs using Space Telescope Imaging Spectrograph (STIS) on board the Hubble Space Telescope (HST) and they found a temporal variation in the evaporating atmosphere. While evaporation was not detected in the first epoch of observations (April, 2010), in the second epoch (September, 2011), transit absorption depths of $14.4\% \pm 3.6\%$ of high-velocity, blue-shifted planetary material were observed in the Ly-$\alpha$ line.
A detection of an X-ray flare just 8h before the second transit observation supports the idea that the atmosphere of HD189733b went through an enhancement of escape caused by a transient event of the host star \citep{Lecavelier12}.

The X-ray flare 
in HD189733A has increased the X-ray energy flux almost by a factor of 3.7 times compared to the quiescent phase \citep{Lecavelier12}. In the Sun, CMEs are most of the time associated with flaring events \citep{Compagnino2017} and considering the fact that the flare has been observed 8h before the observed planetary transit, it is possible that if a CME occurred just after the flare, it would have arrived at the planetary orbit after 8 hours.

In this work, we investigate the interaction of a flare, a CME, and both simultaneously on the atmosphere of HD189733b to find how they are affecting the mass loss rate of the planet and corresponding transit signatures in the Ly-$\alpha$ line. For that, we have developed a 3D hydrodynamic escape model where the stellar radiation ionises the planetary neutral hydrogen material and launches the planetary outflow self-consistently, by solving the radiation-hydrodynamic equations. This gives us a unique opportunity to study the effect of stellar radiation, stellar outflow and corresponding physical changes in the escaping atmospheres more realistically than previous studies.

This paper is organized as follows. In the next section, we present our newly developed model.
We then study planetary evaporation by first assuming the XUV radiation from the star in quiescent state (Section~\ref{sec:pw}) without any stellar wind. In Section~\ref{sec:transients}, we model the effect of transient events on the planetary outflow. After presenting a `quiescent' state of the stellar wind, the effects of an individual flare, individual CME and effect of both together are discussed in this section. The synthetic transit signatures are calculated for all cases in the Ly-$\alpha$ line and a detailed comparison with observations are presented in the Section~\ref{sec:transit}. Finally, our conclusions are given in Section~\ref{sec:conclusion}.

\section{3D Radiation Hydrodynamic Escape Model}\label{sec:model}
To model atmospheric escape and the subsequent interaction of the upper atmosphere with the stellar wind, we developed a new user module for the BATS-R-US code \citep{Toth2005}, based on a previous version of our code published in \citet{Carolan2020, Carolan2021}. BATS-R-US has been widely used to study space weather events \citep{Manchester2004, Manchester2005,Toth2007} and solar system objects \citep{Ma2002,Ma2004,Sterenborg2011,Carolan2019}. For our study, we solve the 3D radiation hydrodynamic equations in the rotating frame of the planet (orbital frame of the system), which we assume to be tidally locked (rotation period is given by the orbital period). Compared to the isothermal model adopted in \citet{Carolan2020, Carolan2021}, our new implementation considers photoionisation, collisional ionisation and their corresponding heating and cooling terms in the code. As a result, we are able to launch the planetary outflow self-consistently due to deposition of the stellar radiation energy. This is the first self-consistent implementation of photoevaporation in the BATS-R-US framework for an exoplanetary system.

\begin{figure}
    \centering
    \includegraphics[width=0.5\textwidth]{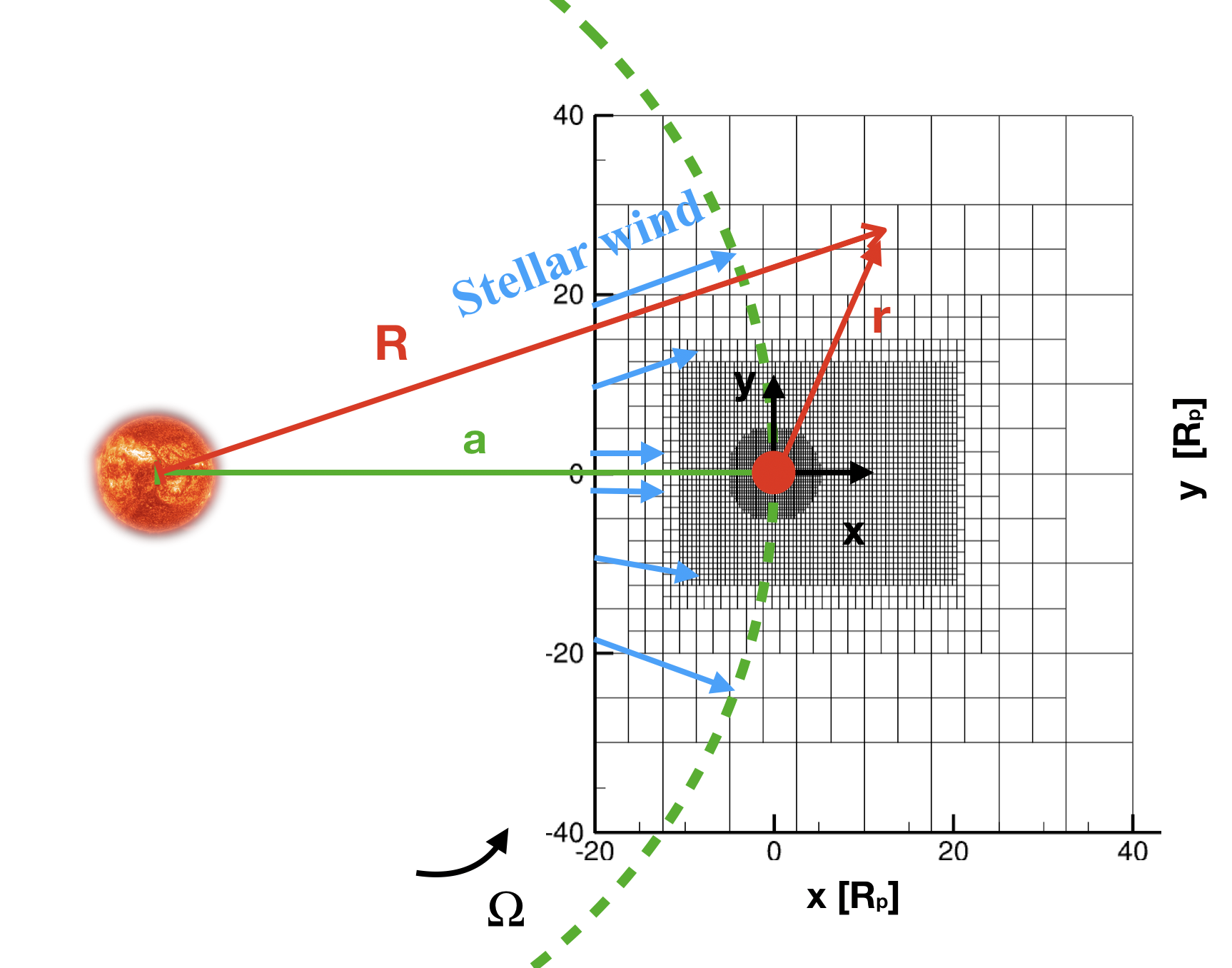}
    \caption{Schematic diagram of our simulation setup. The planet is orbiting the host star in the dashed green track with angular velocity $\Omega$. The grid of our simulation is centered around the planet shown in the grey mesh structure. Stellar wind that is injected from the left side is shown using the blue arrows. }
    \label{fig:grid}
\end{figure}

Our simulations are performed in a 3D Cartesian box keeping the planet at the origin ($x=0,y=0,z=0$, see figure \ref{fig:grid}) and the planetary atmosphere is fully composed of hydrogen. We solve the following mass conservation, momentum conservation and energy conservation equations:
\begin{equation}
\frac{\partial{\rho}}{\partial{t}} +\nabla \cdot \rho {\vec{ u}} = 0,
\end{equation}
\begin{equation}
\frac{\partial(\rho\vec{u})}{\partial t} + \nabla \cdot [\rho \vec{u} \vec{u} + P_TI] = \rho\bigg( \vec{g} -\frac{GM_{*}}{(r-a)^2} \hat{R} - \vec{\Omega} \times (\vec{\Omega}\times\vec{R})-2(\vec{\Omega} \times \vec{u}) \bigg),
\end{equation}
%
\begin{eqnarray}
\frac{\partial }{\partial t} \left( \frac{\rho u^2}{2} + \frac{P_T}{\gamma -1} \right) + \nabla \cdot [\vec{u}(\frac{\rho u^2}{2} + \frac{\gamma P_T}{\gamma -1} )] = ~\nonumber \\ \rho \bigg( \vec{g} -\frac{GM_{*}}{(r-a)^2} \hat{R} - \vec{\Omega} \times (\vec{\Omega}\times\vec{R}) \bigg) \cdot \vec{u}
+{\cal H}-{\cal C},
\end{eqnarray}
where $\rho$, $\vec{u}$ and $P_T$ are the mass density,  velocity and thermal pressure of the material, which is assumed to be composed of hydrogen (neutral and ionised) and electrons,  $I$ is the identity matrix and $\gamma = 5/3 $ is the adiabatic index.  M$_*$ is the mass of the host star. The acceleration due to planetary gravity is $\vec{g} = -GM_p\hat{r}/r^2$, where $M_p$ is the mass of the planet, and $a$ is the orbital separation between planet and star. $\vec{R}$ and $\vec{r}$ are the positional vectors in the stellar and planetary frame of reference respectively. $\Omega$ is the orbital velocity of the planet. Since we solve our system of equations in the rotating frame of the planet, we incorporate the centrifugal force ($\vec{\Omega} \times (\vec{\Omega}\times\vec{R}$)) and the Coriolis force ($2(\vec{\Omega} \times \vec{u})$) in the momentum equation. 
The work done by the centrifugal force is added in the energy equation.

The volumetric heating rate due to the stellar radiation is
\begin{equation}\label{eq:heat}
   {\cal H} = \eta \sigma n_n F_{\rm xuv}e^{-\tau},
\end{equation}
where $\tau$ is the optical depth, $F_{\rm xuv}$ is the XUV energy flux at the orbital distance of the planet and $n_n$ is the number density of neutral hydrogen. The scattering cross-section of the hydrogen atom is
$$
\sigma = 6.538 \times 10^{-32} \left(\frac{29.62}{\sqrt{\lambda}} +1\right)^{-2.963}(\lambda - 28846.9)^2\lambda^{2.0185},
$$
where $\sigma$ is given in cm$^{-2}$ and wavelength $\lambda$ in \AA\ \citep{Verner96, Bzowski13}, and $\eta = (h\nu - 13.6 {\rm eV})/h\nu$ is the excess energy fraction released to heat the gas after a hydrogen atom is ionised (i.e., above the 13.6 eV ionisation threshold). We  assume that the incident stellar radiation is plane parallel and the full XUV part of the spectrum is concentrated at a  monochromatic wavelength $\lambda$ with energy $h\nu$ =  20 eV, which gives us $\sigma$= $1.89\times10^{-18}$ cm$^{-2}$  and $\eta$= 0.32 \citep{Murray-Clay2009,Allan2019,Hazra2020}. We keep the value of F$_{\rm xuv}$ at the left edge of the simulation grid and the flux (F$_{\rm xuv}e^{-\tau}$) which is responsible to ionise planetary material depends on the optical depth $\tau$ of planetary material.
The incident energy flux comes from the negative $x$ direction, which allows us to compute the optical depth at any distance $x$ from the left edge of the grid in the following way
\begin{equation}
    \tau(x) = \int_{x_{\rm left}}^{x}n_n \sigma dx
\end{equation}
where x$_{\rm left}$ is the extreme edge of the grid in our simulation domain.

The total volumetric cooling rate ${\cal C}$ considered here is the sum of cooling due to emission of Ly-$\alpha$ radiation \citep{Osterbrock1989}
$$ {\cal C}_{{\rm Ly}\alpha}= 7.5 \times 10^{-19} n_p n_n \exp{(-1.183 \times 10^5/T )}$$
and collisional ionisation \citep{Black1981}
$$ {\cal C}_{\rm col} = 5.83 \times 10^{-11} n_e n_n   \sqrt{T} \exp{({-1.578\times 10^5}/{T})} \chi_H$$
where T is the temperature, $\chi_H = 13.6 $eV is the ionisation potential of hydrogen, and $n_p$ and $n_e$ are the number density of  protons and electrons, which are the same in a purely hydrogen plasma.  The equations above use cgs units, hence number densities are given in cm$^{-3}$, $T$ in K, and the volumetric cooling rates are in erg~s$^{-1}$~cm$^{-3}$. We assume an ideal gas law, $P_T = n k_B T$, where $k_B$ is the Boltzmann constant and $n = n_e +n_p + n_n$ is the total number density.

In our model, we solve two more equations for tracking the neutrals and ions
\begin{equation}
   \frac{\partial n_n}{\partial t} + \nabla \cdot n_n\vec{u} = \mathscr{R} - \mathscr{I},
\end{equation}
\begin{equation}
   \frac{\partial n_p}{\partial t} + \nabla \cdot n_p\vec{u} = \mathscr{I} - \mathscr{R},
\end{equation}
where $\mathscr{I}$ incorporates the ionisation rate due to photoionisation  and collisional ionisation
\begin{equation}
 \mathscr{I} =\frac{ \sigma n_n F_{\rm xuv} e^{-\tau}}{{h\nu}} + 5.83 \times 10^{-11} n_e n_n \sqrt{T} \exp{({-1.578\times 10^5}/{T})}
\end{equation}
and the recombination rate is
\begin{equation}
 \mathscr{R} = 2.7 \times 10^{-13} (10^4/T)^{0.9} n_e n_p,
\end{equation}
where $\mathscr{I}$ and $\mathscr{R}$ are given in cm$^{-3}$~s$^{-1}$.

In order to solve all the above equations, we need to provide initial conditions and boundary conditions for each fundamental quantity. It is quite common to assume that initially, the planet has a  hydrostatic atmosphere \citep{Tripathi2015, Carroll-Nellenback2017, McCann19, Debrecht2020} and all cells of the simulation box are filled with a hydrostatic atmosphere (i.e., they have a certain value of the density but a null velocity). However, in our case, we initialise our simulation with the steady state planetary outflow from a 1D model as described in \citet{Allan2019, Hazra2020}. The advantage of doing this over an initial hydrostatic atmosphere is that our simulations take less time to reach  steady state.

Our 3D Cartesian simulation box considered here has a rectangular grid with $x = [-20,+40]R_{\rm p}$, $y = [-40,+40]R_{\rm p}$ and $z = [-32,+32]R_{\rm p}$, and the orbital plane is in the $xy$-plane, with the planetary Keplerian motion pointing in the positive $y$ (i.e., the orbital spin axis along positive $z$), as shown in figure \ref{fig:grid}. We use a maximum resolution of $1/16R_{\rm p}$ up to radius of $5R_{\rm p}$  and then $1/8R_{\rm p}$ for $x = [-10,20]R_{\rm p}$, $y = [-12,12]R_{\rm p}$ and $z = [-12,12]R_{\rm p}$ and it reduces further out in the grid (see figure~\ref{fig:grid}).
Note that as the host star is very close to the planet, at a distance of $58.3R_{\rm p}$ from the planet, we keep our negative $x$ boundary shorter (at $-20R_{\rm p}$) than the right boundary ($+40R_{\rm p}$).


We impose an inner boundary at the surface of the planet ($r = 1R_{\rm p}$), where the velocity of the outflow is set to be reflective in the non-inertial frame. 
The base density and temperature are fixed at a constant value. We take a temperature at the base of the planetary atmosphere of $1000$~K, which is approximately the equilibrium temperature of the planet. We use a base density of $4.0 \times 10^{-13}$~g~cm$^{-3}$. The base density of hot Jupiter is not known. We make our decision based on 1D studies \citep{Murray-Clay2009, Allan2019} who showed that, provided that the base of the planetary outflow is in the optically thick region ($\tau > 1$), the flow solution is relatively insensitive to the choice of base density.
At the inner boundary, the base density is mostly dominated by neutral material, with an ionisation fraction set initially to $10^{-5}$.

Similarly to \citet{Carolan2020, Carolan2021}, we inject the stellar wind in our simulation from the negative $x$ boundary. We obtain the physical properties of the stellar wind from a 1D isothermal Parker wind model \citep{Parker1958}. The temperature $T_\star$ and mass-loss rate $\dot{M}$ are free parameters in the Parker wind model, that we choose to set the strength of the stellar wind in terms of the density and velocity. We adopted $T_\star$ and $\dot{M}$ from the stellar wind presented in \citet{Kavanagh2019} for the K dwarf HD189733A. After obtaining the stellar wind solution at each point on the face of negative $x$ boundary, we allow that solution to propagate radially away from the host star into our simulation grid. Once the stellar wind enters the simulation domain, the Coriolis force starts to act on it as we compute all quantities in the rotating frame of the planet. Also since the CME properties are not known currently, we assume it as a denser and faster stellar wind \citep{Cranmer2017}.

The outer boundary conditions are set as an inflow limiting boundary condition. This means that the velocities and other quantities are set to float in the boundary when velocity is directed outwards (i.e., leaving the grid), but if the velocity is directed inwards, the momenta are set to zero. Implementing this kind of boundary condition is necessary because the Coriolis force can bend the flow material near the boundary giving rise to  uncontrolled inflows, which can cause numerical issues if using the usual floating boundary conditions \citep{McCann19, Carolan2020}.

\section{Planetary outflow without stellar wind/CME}\label{sec:pw}
We implement our newly developed model (as described in section~\ref{sec:model}) to the star-planet system HD189733 for studying the planetary outflow and its interaction with
the quiescent stellar wind and stellar transients from the host star. HD189733 is a double star system with a K0V star HD189733A and a M4V companion HD189733B. The planet HD189733b orbits the K star at a distance $a=0.031 {\rm au} = 58.3R_{\rm p}$ and has a mass of $M_p = 1.142~M_{\rm Jup}$ and radius $R_{\rm p} = 1.138~R_{\rm Jup}$ \citep{Llama2013, Stassun2017}.

The host star has a mass $M_\star = 0.82 M_\odot$ \citep{Nordstrom2004}, radius $R_\star = 0.81R_\odot$ \citep{Baines2008} and a rotation period of 12 days \citep{Fares10}.
The star is  about 1-2 Gyr-old and relatively active \citep{Sanz-Forcada11, Poppenhaeger14}.
The measured X-ray luminosity in the quiescent phase of the star is $1.6 \times 10^{28}$ erg~s$^{-1}$ \citep{Lecavelier12}. We calculate the XUV luminosity using the empirical relations given in equation~3 of \citet{Sanz-Forcada11}. The estimated XUV luminosity during the quiescent phase of the star is then $L_{\rm xuv}$  = 1.3 $\times$ 10$^{29}$ erg~s$^{-1}$ , which gives an XUV flux at the orbital distance F$_{\rm xuv}$ = 4.84 $\times$ 10$^{4}$  {\rm erg}~ {\rm cm}$^{-2}$~{\rm s}$^{-1}$.
The initial neutral number density and ionisation fraction at the base of the planet are set to 2.39 $\times$ 10$^{11}$ cm$^{-3}$ and 10$^{-5}$ respectively. Note that our final steady state solution is relatively insensitive to the initial conditions. A change of factor 10 in base density leads to a change of factor 2 in the mass-loss rate. However, a change in base temperature by a factor of 2, does not change the mass-loss rate significantly. 
We keep the same initial conditions for all of our simulations.

\begin{figure}
    \centering
     {\includegraphics[width=0.45\textwidth]{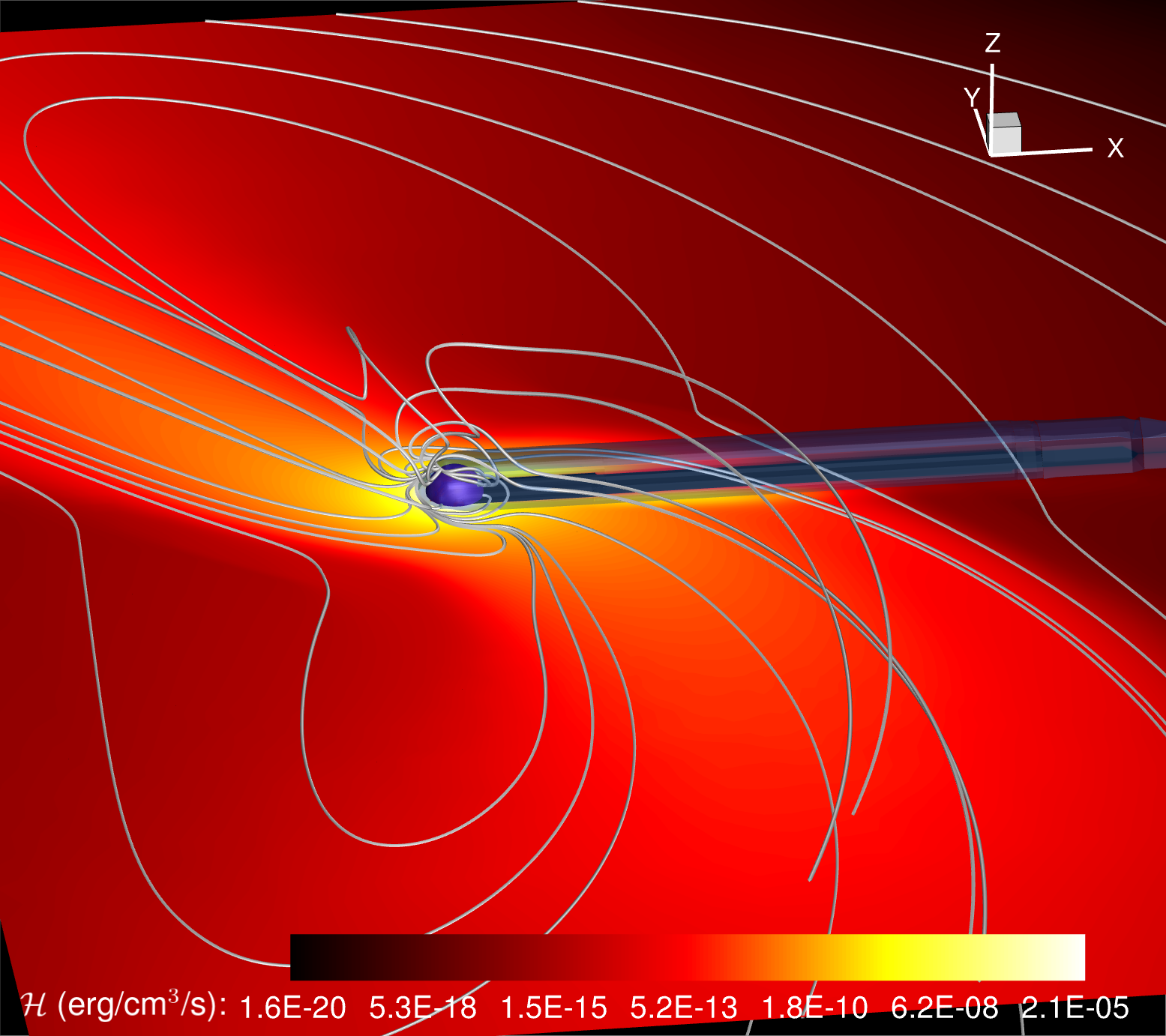}}
    \caption{The volumetric heating profile is shown in the orbital plane of the planet ($z=0$). The grey streamlines show the velocity of the planetary outflow inside our 3D box. The planet shadow is prominent at the night side of the planet. The solid purple color shows the planet surface. The 3D optical depth iso-surface with $\tau = 1$ is shown in transparent blue.}
    \label{fig:heating}
\end{figure}

The heating profile (equation~\ref{eq:heat}) due to incident stellar radiation, which drives the planetary outflow, is shown in figure~\ref{fig:heating}. We show heating only in the orbital plane. The radiation is coming from the negative $x$ direction where the host star is situated and is deposited near the planetary surface where the optical depth is greater or equal to one. The stellar heating after getting deposited around the planetary material shows a swirling nature around the planet due to effect of the Coriolis force. The $\tau =1$ surface is shown by a transparent blue surface, with $\tau$ being greater than one inside that iso-surface. The $\tau = 1$ surface lies just above the planetary surface, where significant amount of photons are deposited. The night side of the planet is not exposed to the stellar radiation and remains cold. As the day and night side temperatures of the planet have a difference, the planetary material is advected to the night side due to pressure gradient. This advected wind gets affected by the Coriolis force as well. The grey streamlines show the velocity of the planetary outflow.

\begin{figure*}
    \centering
    \includegraphics[width=1.0\textwidth]{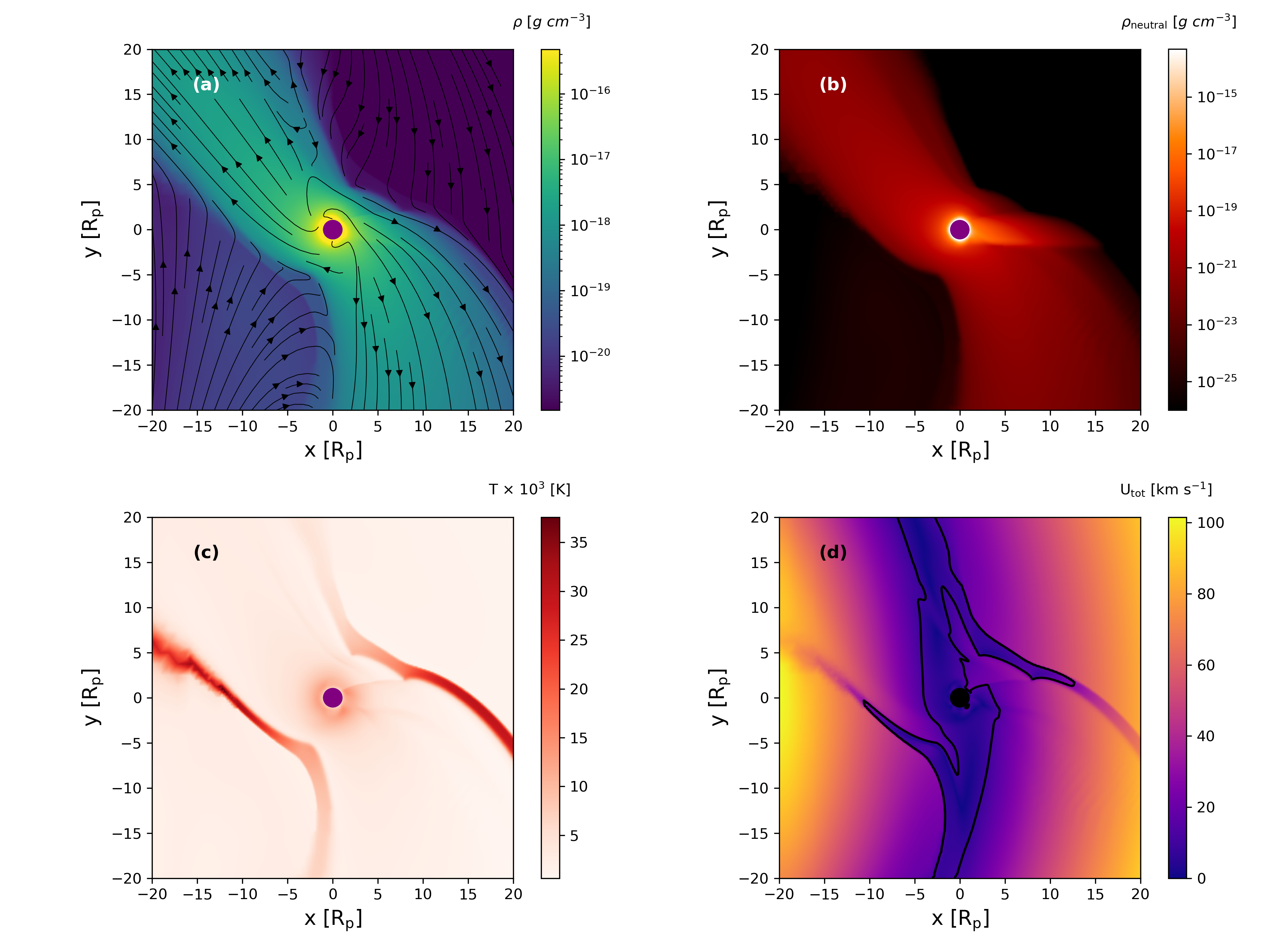}
    \caption{(a) Total mass density along with the velocity streamlines in black. (b) Density of neutral material. (c) Temperature distribution, and (d) Total velocity of the planetary material. Black contour shows the sonic surface where the mach number is one. All plots are in the orbital plane of the planet. The planet is shown as the solid purple circle in each panel. The star is located at the left side of the grid (outside of simulation domain at negative x axis).}
    \label{fig:pw}
\end{figure*}

The basic features of the planetary outflow due to self-consistent deposition of the stellar radiation is shown in figure~\ref{fig:pw}. The total density (sum of ions, electrons and neutrals) of the planetary outflow  is shown in the figure~\ref{fig:pw}(a), where the velocity is shown using black streamlines. The planetary material moves in a clockwise direction after it escapes from the gravity of the planet due to the Coriolis force. However, one thing to note that in spite of a strikingly strong asymmetry in the day and night heating profile, total density is symmetric  across day and night side of the planet. This is because the day-side planetary material gets advected to the night side due to the Coriolis force. The tidal force due to stellar gravity pulls the radially outward planetary outflow after it crosses the Hill radius towards the star but the Coriolis force acts on this material (which is funnelling towards the star) in a clockwise direction. As a result, materials in the quadrant I ($+x,+y$) and quadrant III ($-x,-y$) of our simulation domain get bent by the Coriolis force in the same direction as the tidal force whereas materials in the quadrant II ($-x,y$) and IV ($x,-y$) are pulled  opposite to their Coriolis deflection. After the interplay between tidal force and Coriolis force, the planetary material density is mostly confined in the quadrant II and quadrant IV  as shown in figure~\ref{fig:pw}(a). A shock is formed when the oppositely deflected material meet each other (e.g., the tangential regions between quadrant I \& IV and quadrant II \& III) giving rise to the high temperature in the planetary material as shown in the temperature plot (figure~{\ref{fig:pw}}(c)).

The density of neutral material is shown in figure~\ref{fig:pw}(b). The night side material is no longer ionised when in the planet's shadow and this leads to the formation of a planetary tail of neutral material (figure~\ref{fig:pw}(b)). The material advected here from the day side stagnates until it overcomes the planet's gravity. This neutral tail becomes more bent further from the planet. The total velocity of the planetary outflow is shown in figure~\ref{fig:pw}(d). The black contour is the sonic surface (Mach number $M=1$) of the planetary outflow. All these basic features of our planetary outflow is in accordance with the previous studies \citep{Tripathi2015, Carroll-Nellenback2017, McCann19, Debrecht2020}. We have calculated the mass-loss rate by integrating mass flux through spheres of different radius around the planet. The mass-loss rate from the planet without considering stellar wind is 6.0 $\times$ 10$^{10}$ g~s$^{-1}$. This number will be compared to the cases where we inject the stellar outflow (wind or CME) in the simulation domain (see next Section).

Our estimated planetary mass-loss rate of 6.0 $\times$ 10$^{10}$ g~s$^{-1}$ with F$_{\rm xuv}$ = 4.8 $\times$ 10$^{4}$  erg~ {cm}$^{-2}$~{s}$^{-1}$ is comparable to several existing models for this system.  For example, in previous 1D planetary atmospheric escape models, \citet{Guo2011} found rates of $\dot{M_p}$ = 4.8 $\times$ 10$^{10}$ -- 2 $\times$ 10$^{11}$ g~s$^{-1}$ for F$_{\rm xuv}$ = 2 $\times$ $10^4 - 10^5$ erg~cm$^{-2}$s$^{-1}$; \citet{Guo2016} found $\dot{M_p}$ = 4.5 -- 9.0 $\times$ 10$^{11}$ g~s$^{-1}$ for F$_{\rm xuv}$ = 24778 erg~cm$^{-2}$~s$^{-1}$ for different spectral energy distributions; and  \citet{Salz2016} obtained $\dot{M_p}$ = 1.64 $\times$ 10$^{10}$ g~s$^{-1}$ for F$_{\rm xuv}$ = 20893 erg~cm$^{-2}$~s$^{-1}$. 
Finally, using the energy-limited approximation, \citet{Lecavelier12} estimated a rate of 4.4 $\times$ 10$^{11}$ g s$^{-1}$ during the flaring stage of the host star (100\% heating efficiency). Although the implementation of various physical processes vary among these models, these values are all in reasonable agreement.

\section{Impact of stellar transients on the planetary atmosphere}\label{sec:transients}
\begin{figure*}
     \centering
     \includegraphics[width=0.95\textwidth]{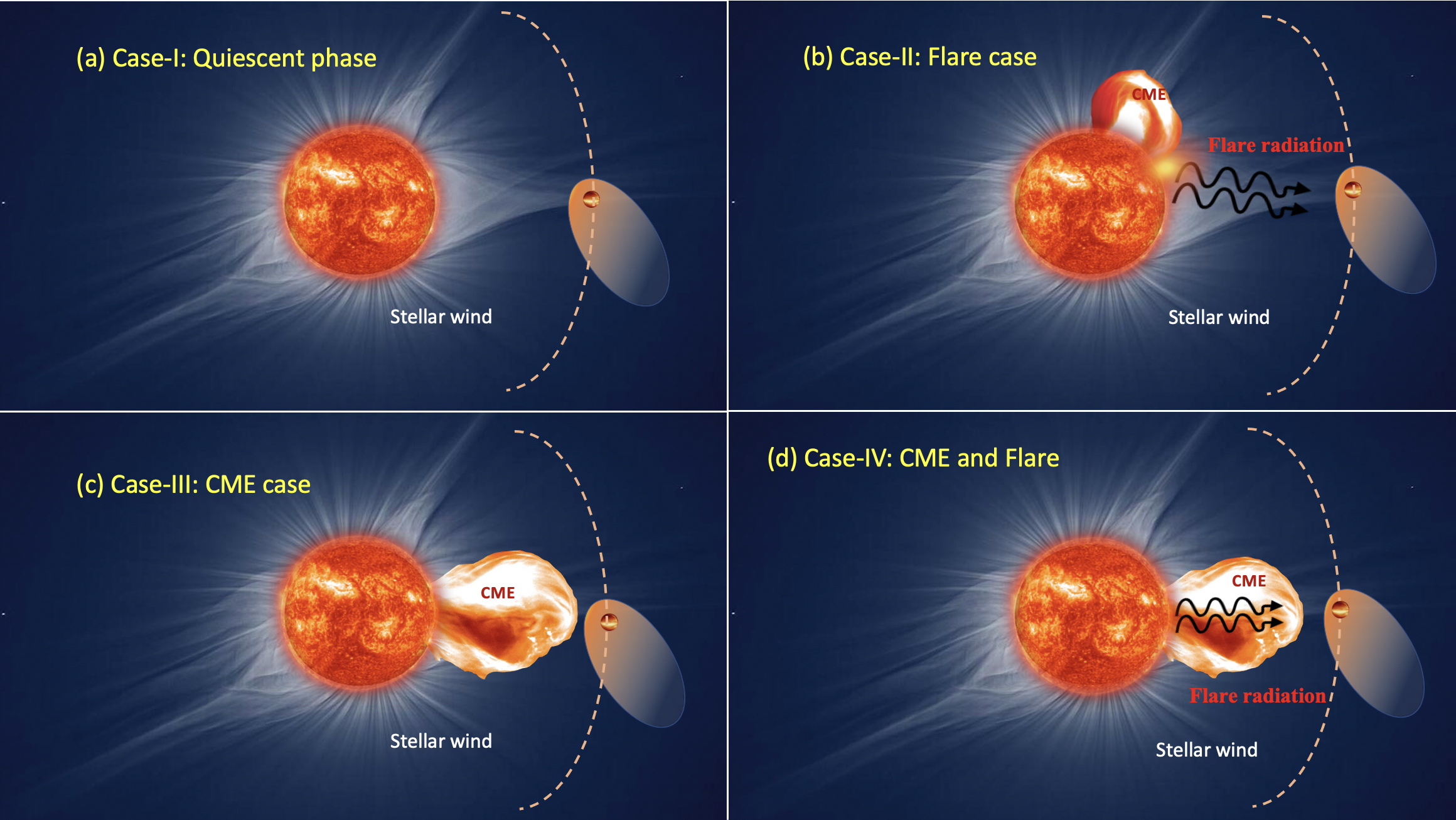}
     \caption{Schematic diagram of four different cases considered in our work. 
     Our planet is orbiting the host star in the dashed orange trajectory. The flare radiation is shown as black arrows and CMEs are shown with enhanced orange material. (a) Case-I: Quiescent phase -- we consider a quiescent background stellar wind and a quiescent XUV radiation. (b) Case-II: Flare case -- a flare is considered by enhancing the XUV radiation that reaches the planet. By not incorporating the effects of a CME, we are assuming that the CME is either not directed to the planet or this could represent a case where a CME was not ejected after the flare. (c) Case-III: CME case -- A CME is directed to the planet, but no flare happens. (d) Case-IV: CME and flare -- The flare and the CME are happening simultaneously. }
     \label{fig:four_cases}
 \end{figure*}

The host star HD189733A is an active star with a rotation period of 12 days \citep{Fares10}. Solar-like stars that rotate faster than the Sun flare more frequently \citep{Maehara2015}. Given that statistical studies of solar CMEs show that CMEs are most of the time associated with flares \citep{Compagnino2017}, to study the effect of stellar transient events on the planetary atmosphere, we consider three instances of transient events, namely: only a flare; only a CME; and a flare associated with a CME. Mostly, flares enhance the stellar radiation and CME enhances the stellar wind condition after their occurrence. Hence, we changed our stellar irradiation and stellar wind parameters according to the case we study. The stellar environments which we want to capture in our model are schematically depicted in figure~\ref{fig:four_cases}:
 \begin{itemize}
     \item {\bf Case-I (panel a)} --
 When the host star is in quiescent phase i.e., there are no flares or CMEs but the (quiescent) background stellar wind interacts with the (quiescent) XUV-driven planetary outflow (see Section~\ref{sec:pw} and figure~\ref{fig:pw} for planetary outflow properties without the stellar wind).
 \item {\bf Case-II (panel b)} --
 The flare case illustrates a situation where the CME is not directed to the planet, hence the planets only ``sees'' an enhancement in the XUV flux. This situation could also represent the case where a flare happens without an associated CME. Note that, in this case, the background stellar wind is still interacting with the planetary outflow but, due to the flare event, the planet will receive more XUV radiation, which modifies the strength of the planetary outflow.
 \item {\bf Case-III (panel c)} --
 Considering the fact that sometimes solar CMEs are observed without an underlying flare \citep{D'Huys2014}, we study a case with a planet facing a CME alone.
 \item {\bf Case-IV (panel d)} --
 Finally, we also study a case where a CME, associated with a flare, is directed towards the planet.
 \end{itemize}

 In Table~\ref{tab:cmeproperties}, we show the parameters we use for different cases. The density and velocity of the stellar wind/CME are the values injected at $\{x,y,z\} = \{-20, 0, 0\}R_p $. With our simulations, we are studying the maximum response of the planetary atmosphere when it meets the stellar wind, flare or CME. That is why, once each simulation for our different cases reaches a relaxed steady state (see Appendix~\ref{sec:apendix1} for details), we consider that final steady state solution for our analysis, i.e., they are snapshots of the event(s) being considered. Table~\ref{tab:cmeproperties} also shows the evaporation rates from the planet. Here, we show the results of the simulations after each run has reached (quasi) steady state.

\begin{table}
	\centering
	\caption{Adopted values of XUV stellar luminosity, mass density of the stellar wind or CME injected into the grid, its speed and the derived evaporation rate of the planet.}
	\label{tab:cmeproperties}
	\begin{tabular}{lccccr} 
		\hline
				Cases & L$_{\rm xuv}$ & density & speed & $\dot{M}_{p}$ \\
		& (10$^{-5}$ L$_\odot$) & (g~cm$^{-3}$) & (km~s$^{-1}$) & (10$^{11}$ g s$^{-1}$) \\
		\hline
		Planetary outflow only &3.4  & -- & -- & 0.6 \\
	I Quiescent phase & 3.4  & 5.3 $\times$ 10$^{-18}$ & 315 & 0.8 \\
    II Flare case & 11  & 5.3 $\times$ 10$^{-18}$ & 315 & 1.0 \\
    III CME case & 3.4  &  2.1 $\times$ 10$^{-17}$ & 755 & 3.2 \\
    IV CME \& Flare & 11  & 2.1 $\times$ 10$^{-17}$ & 755 & 4.0 \\
		\hline
	\end{tabular}
\end{table}

\subsection{Case-I: quiescent phase}
\begin{figure}
    \centering
    \includegraphics[width=0.45\textwidth]{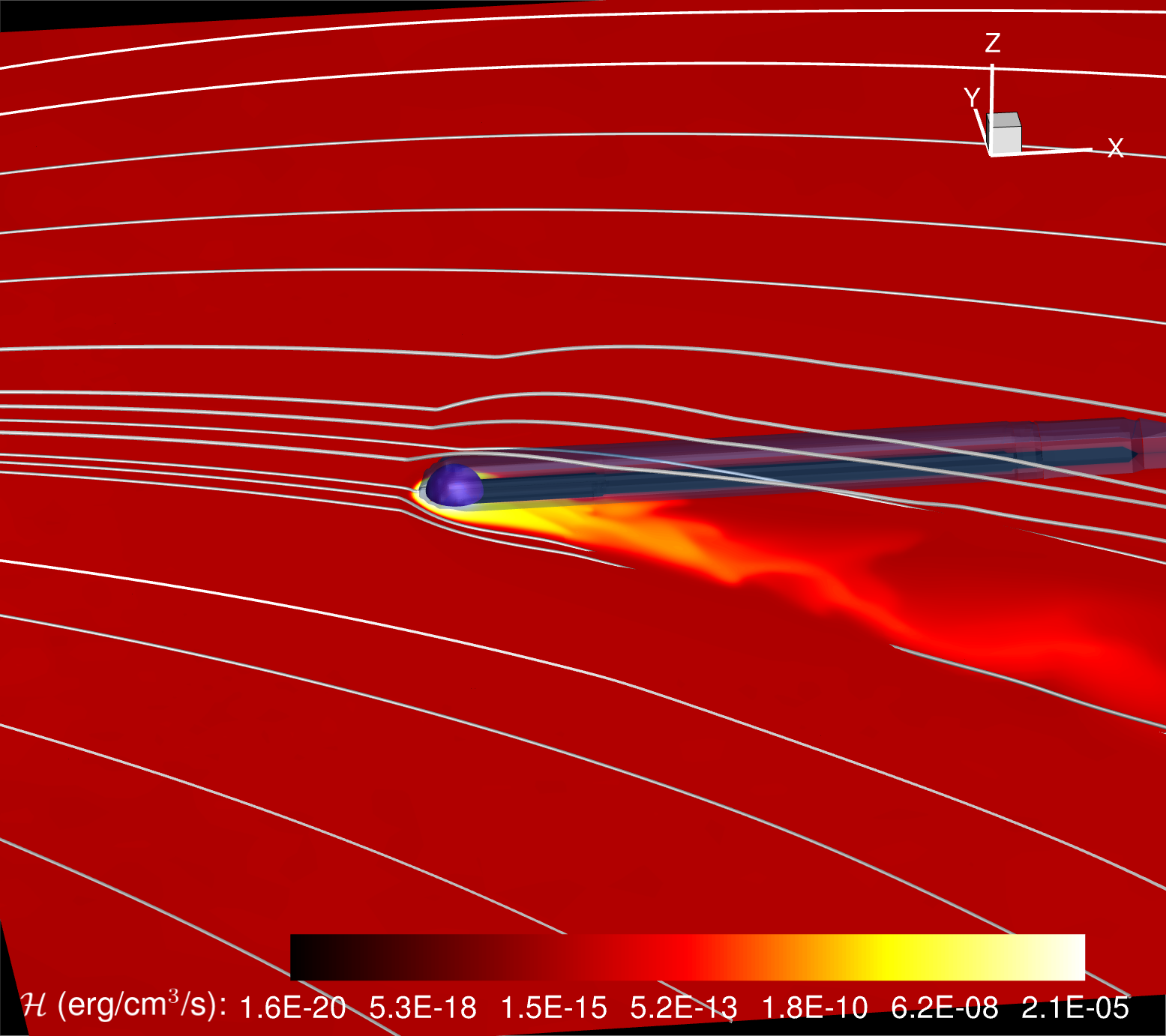}
    \caption{Same as figure~\ref{fig:heating} but for the Case-I with injected stellar wind}
    \label{fig:heating_case1}
\end{figure}

\begin{figure*}
    \centering
    \includegraphics[width=1.0\textwidth]{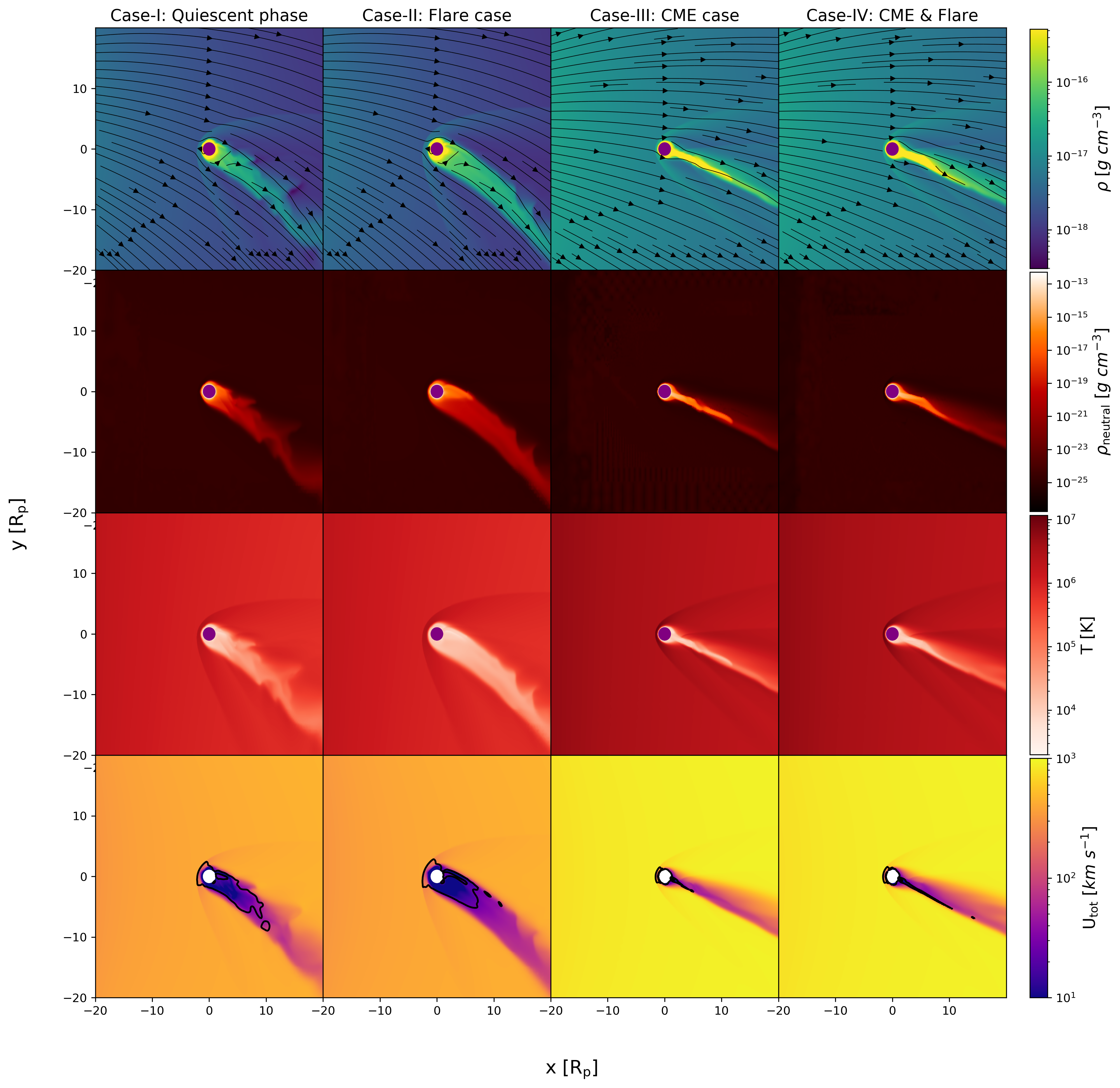}
    \caption{First column: Total density structure, density of neutral material, temperature and total velocity of planetary atmosphere in the orbital plane of the planet for the quiescent phase in the first, second, third and fourth row respectively. Black streamlines in the total density plots are the velocity streamlines. Black contours in the total velocity plots at bottom row shows the sonic surfaces. Second column: Same quantity as first column but for the flare case. Third Column is for the CME case and Fourth column shows the quantities for the CME \& Flare case. The planet is shown in the purple solid circle except in the bottom row which is shown by a solid white circle.}
    \label{fig:transients}
\end{figure*}

The heating distribution due to stellar radiation after inclusion of stellar wind in the quiescent phase of the host star is given in figure~\ref{fig:heating_case1}. The stellar wind pushes the day side material in a tail like structure and heating is mostly confined around the tail. The grey streamlines show the velocity streamlines. The blue transparent surface shows the $\tau=1$ surface. This figure should be directly compared with the heating profile without stellar wind in figure~\ref{fig:heating} to understand how the stellar wind is redistributing the planetary material and hence heating around the planet. In the first column of figure~\ref{fig:transients}, we show the total density distribution, neutral density, temperature and total velocity in the steady state after the stellar wind interacts with the planetary outflow from top to bottom rows respectively for the quiescent phase. The black streamlines in the total density plot shows velocity streamlines. The chosen density of $5.3 \times 10^{-18}$ g~cm$^{-3}$ and speed 315 km~s$^{-1}$ of the stellar wind (at the left edge of our simulation domain) results in a stellar mass-loss rate of $3 \times 10^{-12} \dot{M_\odot}$ yr$^{-1}$. The temperature of the stellar wind is 1.9 $\times$ 10$^6$ K. Our choice of density and speed leads to a high stellar wind ram pressure. As a result, it interacts with the supersonic planetary outflow very close to the planetary surface, as shown in the bottom left of figure~\ref{fig:transients} by the sonic surface in black contour. The mass-loss rate from the planet is $7.6 \times 10^{10}$ g~s$^{-1}$, which is 27$\%$ higher than the mass-loss rate due to planetary outflow without stellar wind. After the interaction with the stellar wind, the planetary material follow the stellar wind streamlines in the planetary tail. This regime is similar to the strong stellar wind mass loss rate case considered by \citet{Carolan2021} [see right panels of figure~5 of their paper].

\subsection{Case-II: flare case}\label{sec:flare}
One of the motivations to choose HD189733 star-planet system is that \citet{Lecavelier12} detected an enhancement in atmospheric evaporation of HD189733b 8h after a flare was observed. This might have been coincidental and given that the link between the two events (evaporation and flare) is not obvious, we want to compare the effect of stellar transients in the atmosphere of the planet. 
The peak X-ray flux observed during the flare event is 1.3 $\times$ 10$^{-12}$ erg~cm$^{-2}$~s$^{-1}$ which gives us a X-ray flux of 21 erg~cm$^{-2}$~s$^{-1}$ at 1 au from the star. Using the empirical relation given in equation~3 of \citet{Sanz-Forcada11}, we calculate the XUV luminosity of the star during the flaring stage. The estimated L$_{\rm XUV}$ = 1.1 $\times$ 10$^{-4}$ L$_\odot$ gives an XUV flux $F_{\rm XUV}= 1.5 \times  10 ^5$ erg~cm$^{-2}$~s$^{-1}$ at the planet orbit.  We use this XUV radiation in our simulation to incorporate the effect a flare. 

The total density, neutral density, temperature, and total velocity distribution for the flare case is shown in the second column of figure~\ref{fig:transients}. A flare induced XUV radiation heats up the planetary atmosphere more than it does during the quiescent phase of the star, which lifts up more planetary material through the outflow and ionises more neutral material. The planetary outflow has higher neutral density (see second row in second column of figure~\ref{fig:transients}) and proton density in the flare case, which eventually leads to a higher planetary mass-loss rate in comparison to the quiescent phase. A denser planetary material results in a stronger ram pressure of the planetary outflow which interacts with the stellar wind higher in the atmosphere than the quiescent phase. The sonic surface for the flare case is higher than the quiescent phase (see the sonic surfaces in the bottom row of second column in figure~\ref{fig:transients}). As a result, the stellar wind can not penetrate very deep into the planetary atmosphere but shapes the planetary material into a cometary tail as in the quiescent stellar wind case (Case-I). The planetary mass-loss rate during this case is $1.0 \times 10^{11}$ g~s$^{-1}$.

\subsection{Case-III: CME only}\label{sec:cme}
In our simulation, we mimic the effect of a CME by changing the properties of the injected stellar wind.
The CME materials are much faster than the quiescent stellar wind. In order to simulate a reasonable CME from HD189733A, we have chosen a density of $2.1 \times 10^{-17}$ g~cm$^{-3}$ and speed of 755 km/s, which are a factor of 4 and 2.4 larger than the quiescent stellar wind, respectively. The temperature of the CME is 5.4 $\times$ 10$^6$ K. For Case-III, we have assumed that there is no flare (i.e., the XUV radiation is the same as the quiescent phase of the star) and the CME is directed towards the planet and hence interacts with the planetary atmosphere.

Once the CME enters the simulation box, it behaves in a similar way as the quiescent phase stellar wind. But in this case, the CME is much denser and faster, which leads to a stronger ram pressure. The strong CME flow can get closer to the planetary surface and almost confine the whole dayside planetary outflow (see third column of figure~\ref{fig:transients}). The sonic surface is closer to the planet in the day side and night side in comparison to the quiescent phase of the stellar wind (black contour in the bottom of row of the third column). This is because for the quiescent stellar wind case, the dayside material could advect to the night side and could stack up at the planet's shadow (unless it crossed the Hill radius at $4.46R_{\rm p}$). However, for the CME case, the night side material is  also swept away by the high velocity CME material.

The planetary tail follows the resultant CME velocity direction after it gets affected in the rotating frame by the orbital motion of the planet. The tail is less affected compared to the stellar wind interacting the planetary outflow (Case-I and Case-II) because the speed of CME is higher than the stellar wind (bottom row of third column in figure~\ref{fig:transients}). Hence, the planetary tail is more aligned towards the star-planet line. The mass-loss rate of this case is 3.2 $\times$ 10$^{11}$ g~s$^{-1}$ which is 5.3 times higher than the planetary outflow only.

\subsection{Case-IV: CME $\&$ flare} \label{sec:cme+flare}
Statistically most of the solar CMEs are associated with a flare \citep{Compagnino2017} and, in this simulation, we consider a case where  a flare was followed by a CME, which is directed towards the planet. 
As the flare (radiation) and CME (particles) arrival times at the planet are different, we assume that the flare has arrived first to the planet. After 
the planetary atmosphere has relaxed, the CME enters the simulation domain mimicking the realism of flare \& CME incidence. We use the same XUV flux as the flare case (section~\ref{sec:flare}) and CME properties as the CME only case (section~\ref{sec:cme}). The overall atmospheric behaviour (fourth column in figure~\ref{fig:transients}) is similar as the CME case only  but the neutral tail is denser than the CME case only (second row in fourth column of figure~\ref{fig:transients}).
The maximum evaporation rate occurs when we consider the CME and flare simultaneously. This is because the CME has the strongest ram pressure among all the cases consider here, which erodes more the planetary atmosphere, while simultaneously the higher XUV flux due to the flare enhances the planetary photoevaporation. The CME $\&$ flare case has a mass-loss rate of 4.0 $\times$ 10$^{11}$ g~s$^{-1}$ which is almost an order of magnitude larger than quiescent phase of the star.


\section{Synthetic transit observation: Ly-$\alpha$ line}\label{sec:transit}
\begin{figure*}
    \centering
    \includegraphics[width=0.9\textwidth]{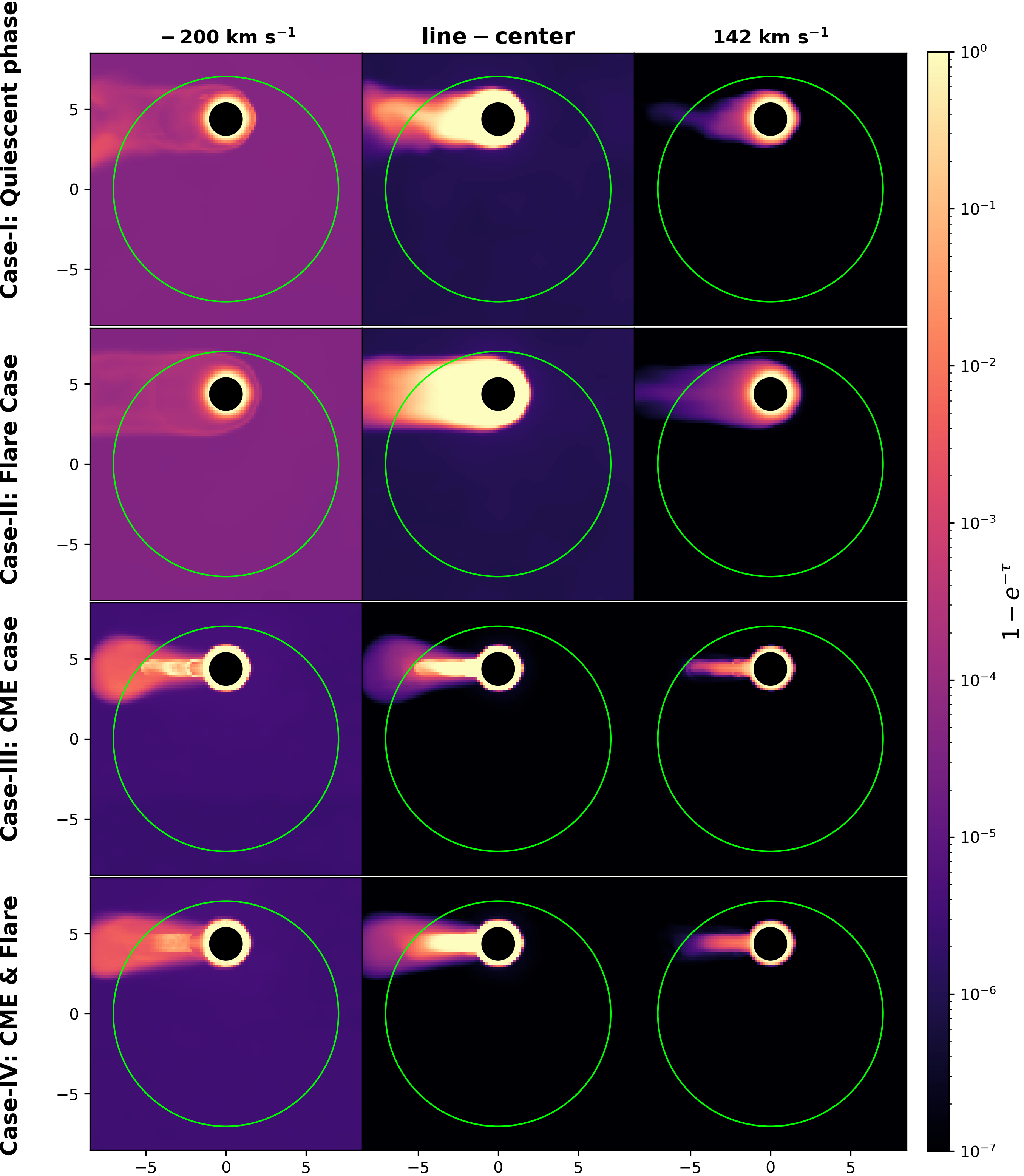}
    \caption{The absorbed specific intensity ($1-e^{-\tau_\nu}$) for blueshifted material at $-200$ km s$^{-1}$, at line center and for redshifted planetary material $142$ km s$^{-1}$ are shown in the left, middle and right panel respectively in the plane of the sky. Four rows from top represent the four cases from Case-I to Case-IV. The green circle is the stellar disc and solid filled black circle shows the planet.}
    \label{fig:absorp}
    \end{figure*}

\begin{figure*}
    \centering
    \includegraphics[width=0.4\textwidth]{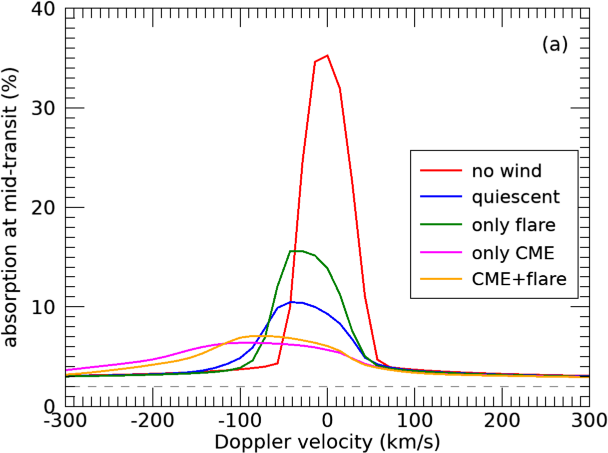}
    \includegraphics[width=0.4\textwidth]{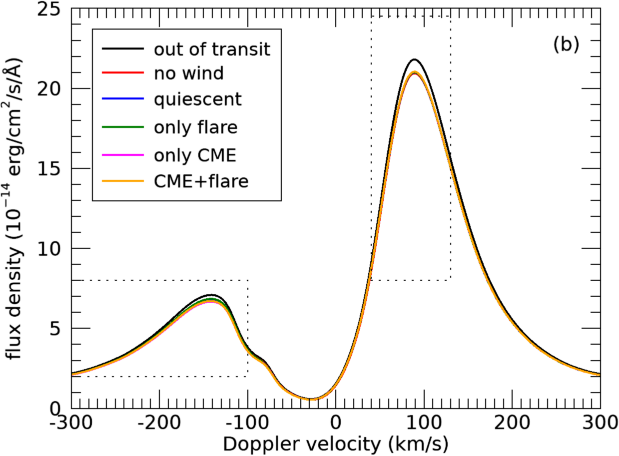}
    \includegraphics[width=0.4\textwidth]{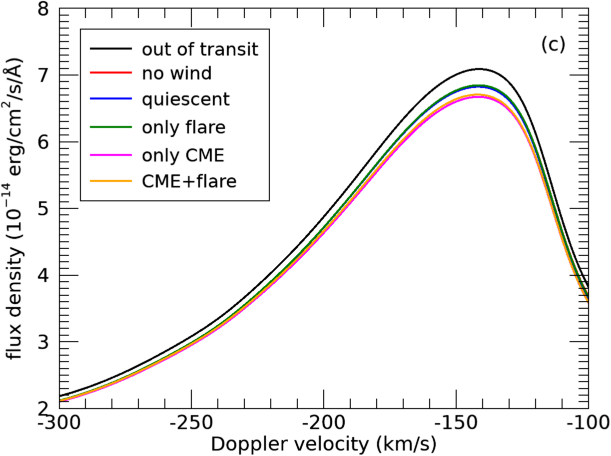}
    \includegraphics[width=0.4\textwidth]{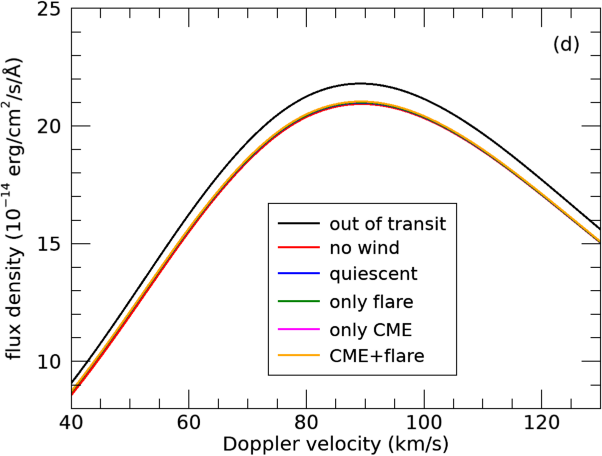}
    \caption{Ly-$\alpha$ line profile computed at mid transit. (a) theoretical line profile; (b) predicted line profile, convolved with the line spread function of the  G140M grating mode; (c) same as b, but zoomed in in the blue wing of the line; (d) same as b, but zoomed in in the red wing of the line.}
    \label{fig:transit_dept}
\end{figure*}

To compare with the available transit observation of HD189733b, we  calculate the transit spectra for all our cases. For that, we use the ray-tracing method presented in  \citet{Vidotto2018b, Allan2019}. As our simulation is performed in the rotating frame of the planet, we convert the velocity to the inertial frame of reference (observer's frame). We then interpolate our 3D simulation output into a regularly spaced 3D Cartesian box, with 251 points in each direction, where the planet is at the center of this box. We use 71 velocity channels from $-500$  to $500$ km s$^{-1}$ to shoot frequency/velocity-dependent rays through the plane of the sky [251 $\times$ 251 cells]. The specific intensity of the radiation after passing through each pixel is $I_{\nu} = I_{\star} e^{-\tau_\nu}$, where I$_{\star}$ is the specific stellar intensity and $\tau_\nu$ is the  frequency/velocity dependent optical depth. The optical depth at each frequency is
\begin{equation}
    \tau_\nu = \int n_n\sigma_t\phi_\nu dz,
\end{equation}
where $\sigma_t$ is the absorption cross section at the line center and $\phi_\nu$ is the Voigt line profile (see section~4.2 in \citet{Hazra2020} for details). $n_n$ is the neutral density that is directly coming from the simulations. Therefore, the fraction of absorbed specific intensity is $(I_\star - I_{\nu})/I_{\star}=1-e^{-\tau_{\nu}}$. To calculate the theoretical transit depth in the synthetic Ly-$\alpha$ spectrum, we integrate the absorbed specific intensity for all rays over all pixels $\{i,j\}$ in the plane of the sky and then divide by the flux of the star
\begin{equation}
    \Delta F_{\nu} = \frac{\int\int(1-e^{-\tau_\nu}){\rm d}i {\rm d}j}{\pi R_\star^2} .
\end{equation}
The absorbed specific intensities for the four cases considered here are shown in figure~\ref{fig:absorp} and the transit depths at mid-transit as a function of Doppler velocities are shown in figure~\ref{fig:transit_dept}(a). The red, blue, green, magenta and orange lines show the planetary wind-only case, Case-I (quiescent), Case-II (flare only), Case-III (CME only) and the Case-IV (CME $\&$ Flare), respectively.

Figure~\ref{fig:absorp} shows a view of the sky plane with the planet at mid-transit with an impact parameter of 0.6631. The green circle represents the stellar disc. The first, second and third columns correspond to the absorbed specific intensities at three different velocities: blueshifted material at $-200$ km s$^{-1}$, at line center and redshifted material at $142$ km s$^{-1}$ respectively. The first row in figure~\ref{fig:absorp} shows the quiescent stellar wind case, where we see that the density of neutral material (hence the absorbed specific intensity) is smaller in the high blueshifted and redshifted velocities as compared to the line-centre (see blue line in figure~\ref{fig:transit_dept}(a)).

The second row of figure~\ref{fig:absorp} shows the flare case. As the total material is accelerated due to an enhanced  XUV flux, the amount of lifted material is
larger (see green line in figure~\ref{fig:transit_dept}(a)). However,  due to more ionisation of neutral material, there is less neutrals in the higher blueshifted velocities, where the ionising photon is able to penetrate.

The CME case
is shown in the third row of figure~\ref{fig:absorp}, which can be compared with the first row which has the same F$_{\rm XUV}$ and hence the same planetary outflow profile. Because the CME is denser and faster than the stellar wind itself, it carries away more planetary material with it while interacting with the planetary outflow. This results in higher blueshifted absorptions. However, because of the stronger ram pressure, the planetary material is forced to occupy a smaller volume, so overall there is less absorption in Case-III as compared to Case-I (compare magenta and blues lines in figure~\ref{fig:transit_dept}(a)).

In the last row of figure~\ref{fig:absorp}, we show the absorption due to the CME $\&$ Flare case.
This is similar to the third row with a slightly smaller absorption for Doppler velocities $<-120$ km~s$^{-1}$ in the blue wing, as can be seen in Figure~\ref{fig:transit_dept}(a).

In summary, the transit depths at line center are higher for the two cases where we considered flares, compared to the quiescent cases for both stellar wind (compare blue and green lines in figure~\ref{fig:transit_dept}(a)) and CME (magenta and orange lines). The stellar CME affects the high velocity neutral tail of the planetary material (mostly blueshifted material) giving rise to  stronger absorption in the blueshifted velocities of the CME cases (Case-III and Case-IV, magenta and orange lines) than the stellar wind cases (Case-I and Case-II, blue and green lines). This simple comparison shows how important the stellar wind, and more specifically the CME, is in producing a blueshifted absorption and an asymmetric profile of the transit spectra.

To compare with the observations, we apply our theoretical absorptions to the reconstructed Ly-$\alpha$ line presented in \citet{BL2013}. In their Figure 4, these authors show both the intrinsic line profile from the star (the flux density), as well as the line profile after it has been absorbed by the neutral ISM material (hydrogen and deuterium). We use the latter one as representing the out-of-transit line profile $F_{{\rm out},\nu}$, hence the in-transit flux density is
\begin{equation}
F_{{\rm in},\nu} = (1 - \Delta F_\nu) F_{{\rm out},\nu}.
\end{equation}
Finally, to compare with the reported observed values, we convolve our line profiles with the G140M grating of STIS. We adopt a double Gaussian for describing the line spread function (LSF) as discussed in \citet{2017A&A...597A..26B}: the Gaussian representing the core of the line has a FWHM of 1.03 STIS pixels, while the Gaussian representing the wings, has a  FWHM of 5.8 STIS pixels. For STIS, the pixel width (dispersion) is 0.053\AA. Finally, we use the peak amplitude ratios between the core and wing to be 0.95. These values correspond to the instrument LSF derived in `visit 3' by \citet{2017A&A...597A..26B}, but we note that choosing other `visit' values had negligible effects in the results reported below.

The convolved lines are shown in figure~\ref{fig:transit_dept}(b), where we also show the out of transit Ly-$\alpha$ line (absorbed by the ISM and convolved with the instrument LSF) in black. The coloured lines represent each of the cases studied here after the Ly-$\alpha$ line is also absorbed by the planetary atmosphere.  A zoom in of the blue wing ($-230$ to $-140$ km s$^{-1}$) and red wing ($60$ to $110$ km s$^{-1}$) are shown in panels c and d of figure~\ref{fig:transit_dept}. The integrated red wing absorptions are $3.7\%$ and $3.5\%$ for the stellar wind cases (Case-I and Case-II) and CME cases (Case-III and Case-IV) respectively. We note however that the $5.5 \pm 2.7$\% redshifted absorption quoted by \citet{BL2013} could be due to statistical noise in the data, and we do not analyse it further. These line profiles are shown at mid-transit, but to compute the lightcurve we also use different times in the transit.

\begin{figure}
    \centering
    \includegraphics[width=0.4\textwidth]{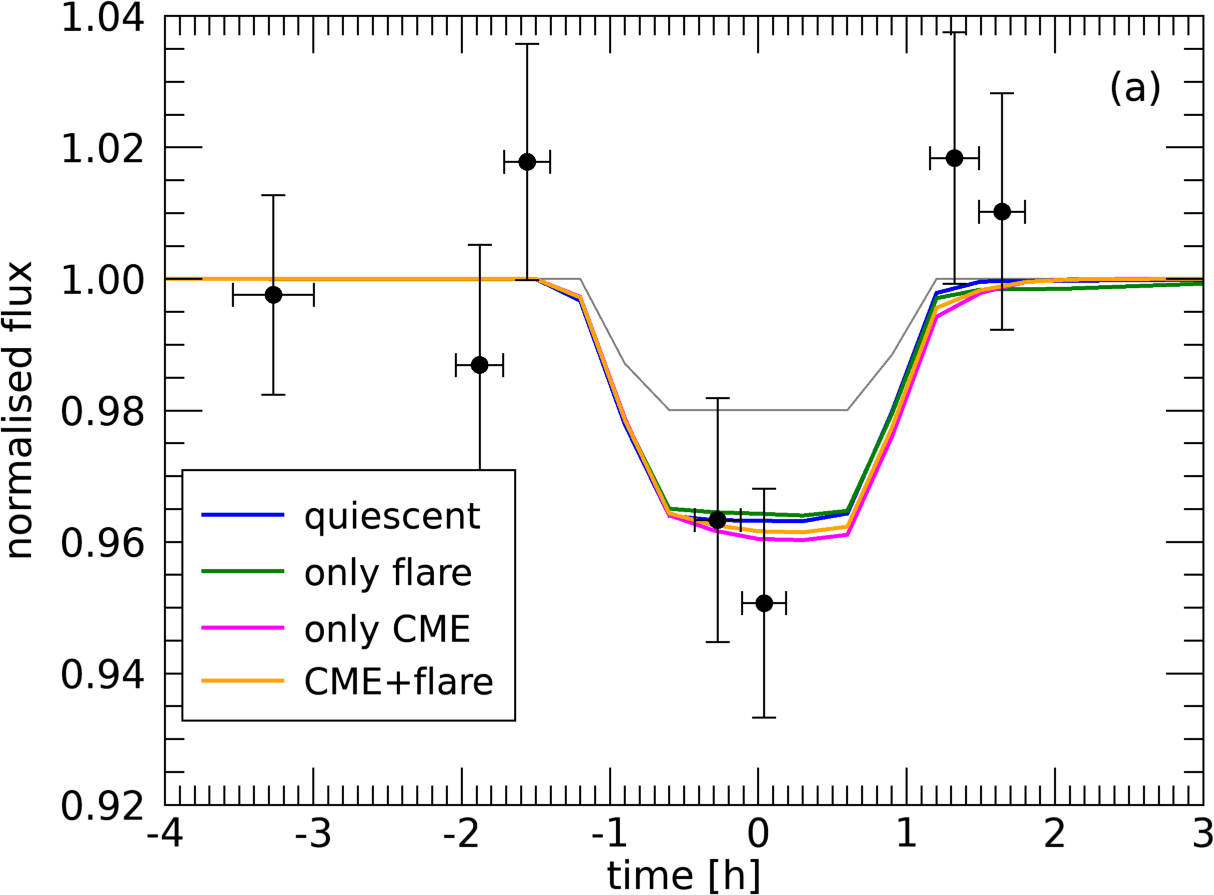}
    \includegraphics[width=0.4\textwidth]{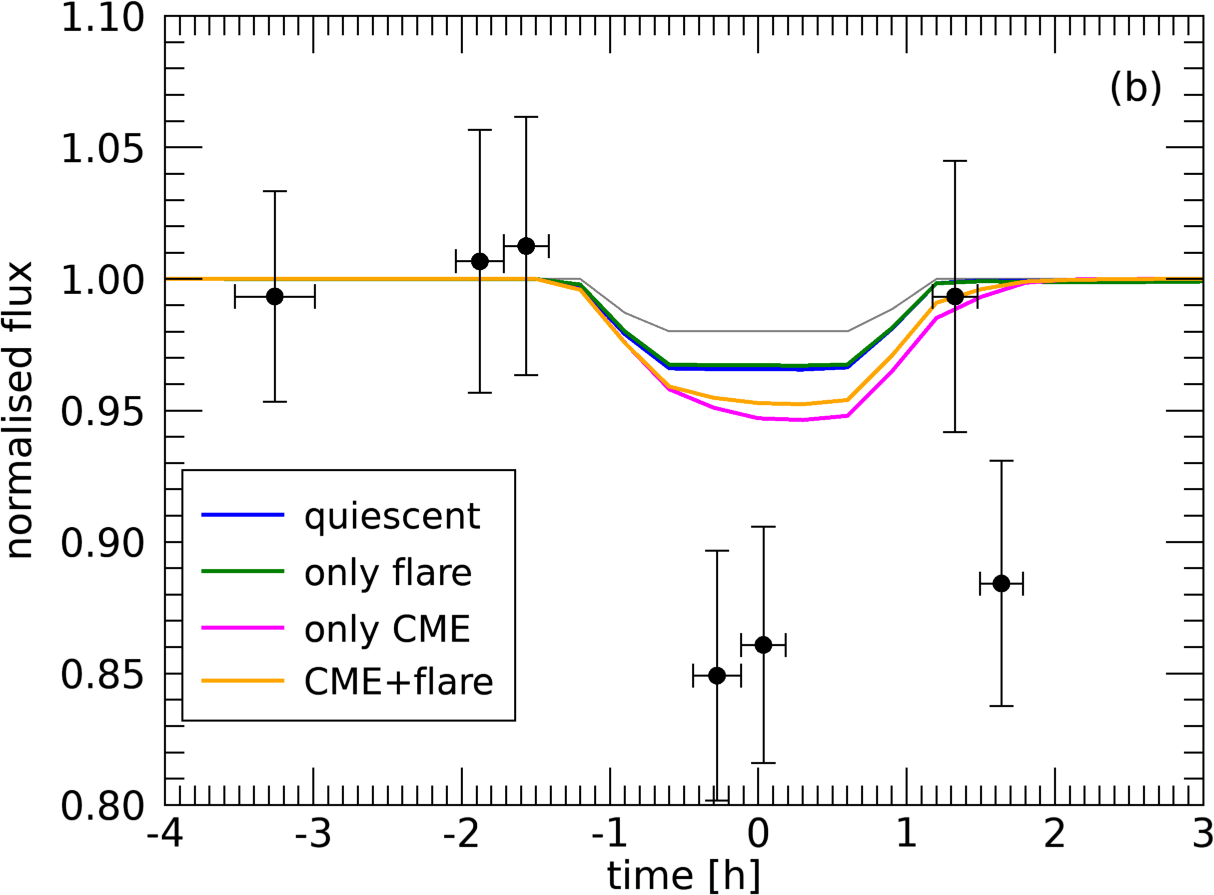}
    \caption{Lightcurves predicted with our models compared to the observed values (circles) from \citet{BL2013} integrated over (a) $-300$ to $300$ km~s$^{-1}$ and (b) $-230$ to $-140$ km~s$^{-1}$ }
    \label{fig:lightcurve}
\end{figure}

Figure \ref{fig:lightcurve}(a) shows the lightcurve of our predicted transits. For that, we integrate our flux densities $F_{{\rm in},\nu}$ and $F_{{\rm out},\nu}$ over the velocity ranging from $-300$ to $300$ km~s$^{-1}$. The circles represent the data shown in Figure 9 of \citet{BL2013}. We note that all the models shown here can represent the observed transit, when considering the total flux of the line. However, the largest absorption is seen in the blue wings of the Ly-$\alpha$ line \citep{Lecavelier12,BL2013}. Panel b shows the lightcurve when the line profiles are integrated only in the blue wing, i.e., from $-230$ to $-140$ km s$^{-1}$. The circles represent the data shown in Figure 14b of \citet{BL2013}. Here, we see that none of our models are able to explain the large absorption seen in the observations. While the cases where we considered CMEs (both with quiescent XUV flux or the flaring XUV flux) show deeper transit lightcurves, our largest integrated blue wing absorption has a depth of $5.1\%$ (Case-III), while the observations show depths of $14.4 \pm 3.6\%$.

\section{Discussion and Conclusions}\label{sec:conclusion}
In this paper, we have presented 3D radiation hydrodynamic simulations of the planetary outflow from HD189733b and its interaction with stellar transient events. Our simulations included both neutral and ionised hydrogen (multi-species) and considered photoionisation, collisional ionisation and radiative recombination. All simulations were performed in the rotating frame of the planet. The salient features of the planetary outflow including Coriolis and tidal forces are in accordance with previous studies \citep{Tripathi2015, Carroll-Nellenback2017, Villarreal2018, McCann19, Debrecht2020}. After the planetary outflow is self-consistently launched  from the planet's surface, we have considered its interaction with the stellar wind, where the incident F$_{\rm xuv}$ flux is considered here during the quiescent phase of the star (Case-I). We further studied three cases which included the effects of stellar transient events: a flare (Case-II), a CME (Case-III) and a CME $\&$ a flare simultaneously (Case-IV).

During the quiescent stellar wind case (Case-I), the planetary outflow is shaped into a comet-like tail by the stellar wind. The day side outflow is mostly confined by the stellar wind and most of the planetary material escapes through the comet-like tail. The flare case (enhanced F$_{\rm xuv}$ but with the same quiescent phase stellar wind) gives a similar structure as the quiescent stellar wind case. However, the flare initiates more evaporation from the planet, leading to a higher mass-loss rate than in Case-I. The CME case (Case-III) mimicking a much denser and faster stellar wind than Case-I gives rise to a similar planetary tail, but the orientation of the tail is closer to the line that joins star and planet. This is because the direction of the tail is roughly given by the vector sum between the orbital velocity and the stellar wind/CME velocity \citep{2010ApJ...722L.168V}. Thus, the tail orientation changes because the CMEs are faster than stellar wind. The case with CME $\&$ Flare shows more evaporation (higher XUV flux) but the planetary outflow morphology is similar to the CME case alone.

The mass-loss rate for the quiescent phase, flare case, CME case and CME $\&$ Flare case are 0.8 $\times$ 10$^{11}$, 1.0 $\times$ 10$^{11}$, 3.2 $\times$ 10$^{11}$ and 4.0 $\times$ 10$^{11}$ g s$^{-1}$ respectively. The CME cases with and without a flare impacts atmospheric mass-loss rate significantly. If the host star is young and active, the CME occurrence would be more frequent and that could have a stronger contribution in the evolution of the planetary atmosphere. The long term effective planetary mass loss can be roughly estimated. First, we need an estimate of the CME occurrence rate over the time period of the planet's life. Second, we need to consider some specific facts to give an estimate on the planetary evaporation rate to be physically accurate. These are evolution history of the stellar wind mass-loss rate, CME occurrence rate, and evolution history of the stellar radiation. But to get an order of magnitude calculation neglecting those facts and assuming the mass-loss rate of the stellar wind, and stellar radiation remain constant and they are same as our Case-I over 1 Gyr of the planet's life, the planetary mass loss due to the stellar wind alone is $1.33 \times 10^{-3}$ $M_{\rm Jup}$. If we assume that the CME occurrence rate is one per day and CME mass-loss rate is constant over the life of the planet and also, each CME interacts with the planetary atmosphere for one hour (we assume this for calculation purpose and it may vary), then the total planetary mass loss due to CME is = $1.65 \times 10^{-4}$ $M_{\rm Jup}$. Therefore the effective long term mass loss is $1.5 \times 10^{-3}$ $M_{\rm Jup}$.

The effects of CMEs on Earth-like secondary atmospheres in close-in orbit planets have been studied previously by  \citet{Khodachenko2007, Lammer2007}. They found that unmagnetized exoplanets are in danger of being stripped of its whole atmosphere by high density CMEs during a 1-Gyr period. \citet{Cohen2011} investigated the effects of CMEs on HD189733b, finding that the planetary magnetosphere is significantly affected by the CME event. However, the modelling of the planetary atmosphere in their simulation was still preliminary, neglecting a more self-consistent physical process to drive the flow. A more realistic modelling of the planetary atmosphere to study the influence of CMEs on the mass-loss rates of hot Jupiter was carried out by \citet{Cherenkov2017}, albeit their models did not include magnetic fields (like ours). They concluded that the amount of mass-loss from the planet due to CME in 1 Gyr is comparable to that due to high energy stellar flux. Our results indicate that CMEs have an important effect on the planetary mass-loss rate too (the planetary mass loss rate due to CME is almost one order of magnitude higher than the quiescent phase of the star) and they should be taken into account in the evolution of planetary atmosphere, in particular at early ages, when CME occurrence is expected to be higher.

We calculated the transit depth in Ly-$\alpha$ line for the four cases to compare with the observed Ly-$\alpha$ transit spectra. \citet{Lecavelier12} and \citet{BL2013} reported a deeper transit in the blue wing of the Ly-$\alpha$ line 8 hours after a stellar flare, raising the question of whether the flare, or an associated CME, led to an increase in the evaporation of the planet. Among all the models considered here,  Case-III (only CME) showed the highest Ly-$\alpha$ absorption. The quiescent and flare cases (Case-I and Case-II) are not able to generate a strong absorption in the blue wing of the Ly-$\alpha$ line. However, CME cases with and without flare (Case-IV and Case-III) show larger absorption of blueshifted material, as compared to Case-I and Case-II, as
the high velocity CME carries away planetary materials with it. Therefore, the observed temporal variation in the blue wing seen in the observations \citep{Lecavelier12} is more likely to be a consequence of the CME impacting the planetary atmosphere, than of the increased energy flux caused by flares.

\citet{Odert2020} also modelled evaporation in HD189733b to investigate the observed Ly-$\alpha$ transit variability. Their investigation included, like ours, the increase in XUV flux due to a flare and  different stellar environment conditions caused by a stellar wind/CME. The escape rates they found (2.5 -- 17 $\times$ 10$^{10}$ g~s$^{-1}$) 
are somewhat similar to ours (8 -- 40 $\times$ 10$^{10}$ g~s$^{-1}$). Additionally, they reported no enhancement of the Ly-$\alpha$ absorption flux during a flare in agreement with our findings. The major difference between our results is on the expected transit with stellar wind/CME.
\citet{Odert2020} found that the stellar wind/CME conditions had a  negligible effect on the Ly-$\alpha$ absorption, which is different from our findings (compare their figure 13 with our Figure 8a). We believe our results differ because of our different modelling approaches.   \citet{Odert2020} simulated the atmosphere of the planet using a 1D hydrodyncamical escape model and, separately, they simulated the interacting stellar wind/CME using a 3D model. The inner boundary to their 3D stellar wind interaction model is at the point of pressure balance (between 2 and 2.6~$R_p$), which takes the values obtained from the 1D atmospheric model. Thus, in their simulation setup, the planetary outflow does not react to the interaction of the stellar wind and they are unable to model the formation of a comet-like tail. For example, their planetary escape rate (obtained from their 1D model) is the same regardless of the stellar wind/CME conditions. In our self-consistent approach, escape rates vary in response to the stellar environment, even when we consider the same XUV flux. Compare, e.g., Cases I and III or II and IV in Table~\ref{tab:cmeproperties}.
The presence of a comet-like tail is particularly important when modelling Ly-$\alpha$ lightcurves, as the tail breaks the symmetry in transit profiles (we refer again to our figure 8a and their figure 13).

We have also modelled additional CME cases with different densities and velocities. For example, a CME that is 2 times denser than the CME considered here (Case-III and Case-IV) increases the planetary mass-loss rate 1.6 times without any significant changes in the Ly-$\alpha$ transit. Similarly, a 2.2-times increase in the CME velocity increases the mass-loss rate by 1.4 times but no significant changes in the  Ly-$\alpha$ line. This means that, although the CME properties affect the evaporation rate, the transit signatures are relatively insensitive to our choice of CME parameters.

Even though our CME case (Case-III) showed the highest Ly-$\alpha$ absorption of 5.1$\%$, it still does not reproduce the observed absorption of $14.4\pm 3.6\%$ (our results agree with the observed value within 2.6$\sigma$).
A possible reason for our departure from observation may be attributed to the non-inclusion of charge exchange and radiation pressure. Radiation pressure from Ly-$\alpha$ photons can accelerate neutral material from the planet towards the observer, hence increasing the amount of blueshifted material. However, as discussed in \citet{Villarreal2021}, if the stellar wind/CME ram pressure already provides the bulk of this acceleration, then the contribution from radiation pressure becomes less significant. Charge exchange  is mostly important in the interaction region between stellar  and planetary material, where hot protons from the stellar wind exchange charges with cold neutrals from the planetary atmosphere, producing energetic neutrals. \citet{Khodachenko2019} found that these energetic neutrals are important in enhancing the Ly-$\alpha$ transit depth in the high blueshifted velocities for the GJ436b. \citet{Odert2020}, however, found that charge exchange does not significantly enhance the transit depth in the Ly-$\alpha$ in the case of HD189733b. \citet{Esquivel2019} also reported similar result in the case of HD209458b. Although charge exchange alone may not be helpful to produce enhanced blueshifted transit depth, note that an inclusion of radiation pressure along with charge exchange could help further accelerate the high-velocity neutrals generated by charge exchange, producing an enhancement of transit depth at higher blueshifted velocities \citep{Esquivel2019}.

Another factor that can enhance the density of the high-velocity neutral material is magnetic fields, which were not included in our simulations. Planetary magnetic field plays an important role in confining the outflow materials inside the dead zones depending upon its geometry and strength, where the ram pressure of the outflow is insufficient to overcome magnetic stresses \citep{Trammell2014, Khodachenko2015}. This leads to a higher transit depth than the non-magnetic case (Carolan et al. 2021b, in prep., and see Section~4 of \citet{Trammell2014} for details). \citet{Trammell2014} showed that for planets with strong magnetic field strengths (dipolar magnetic field at pole $B_0>10$ G), the dead zones can trap high velocity neutral material, causing enhanced absorption in the Ly-$\alpha$ line wings. On the other hand,  \citet{Villarreal2018} found that in the case of HD209858b the Ly-$\alpha$ absorption increases when the planetary magnetic field is reduced for $B_0$ smaller than 5G. Inclusion of planetary and stellar magnetic fields will be the subject of a future study. Additionally, the magnetic configuration of the incoming CME could play an important role in the interaction between planetary outflow and CME -- for example, magnetic reconnection could change the topology of the planetary magnetic field, by opening up more field lines, and thus affecting the planetary outflow properties.

\section*{Acknowledgements}
This project has received funding from the European Research Council (ERC) under the European Union's Horizon 2020 research and innovation programme (grant agreement No 817540, ASTROFLOW). The authors wish to acknowledge the SFI/HEA Irish Centre for High-End Computing (ICHEC) for the provision of computational facilities and support. This work used the BATS-R-US tools developed at the University of Michigan Center for Space Environment Modeling and made available through the NASA Community Coordinated Modeling Center. We thank the
referee, Dr. Luca Fossati, for constructive comments. We also thank another anonymous referee for comments that helped to improve our manuscript.

\section*{Data Availability}
The data underlying this article will be shared on reasonable request to the corresponding author.



\bibliographystyle{mnras}
\bibliography{myref_exoplanet,reference} 




\appendix
\section{A note on steady (quasi-steady) state in our simulation}\label{sec:apendix1}

\begin{figure}
  \centering
  \includegraphics[width=0.43\textwidth]{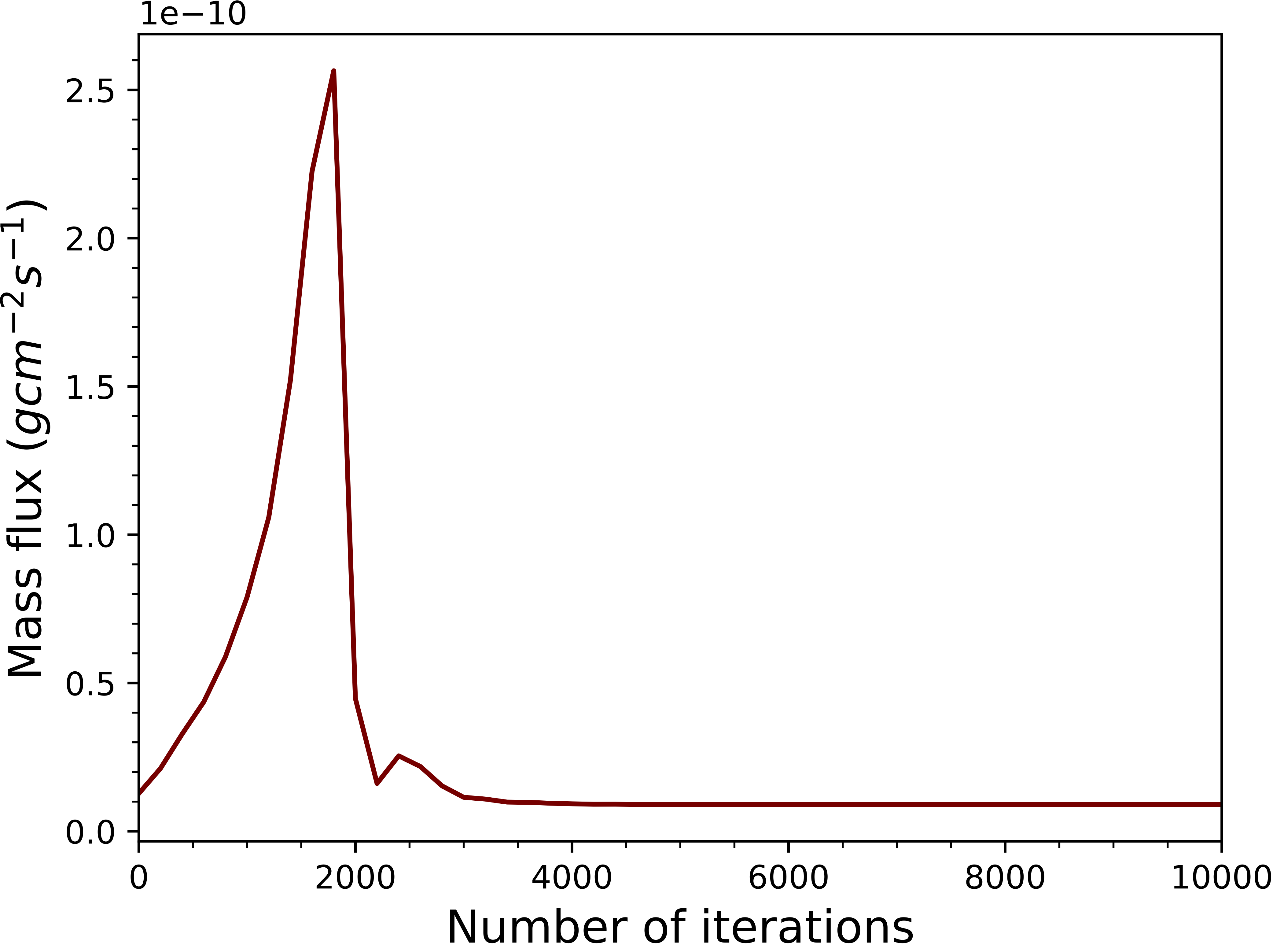}
  \includegraphics[width=0.43\textwidth]{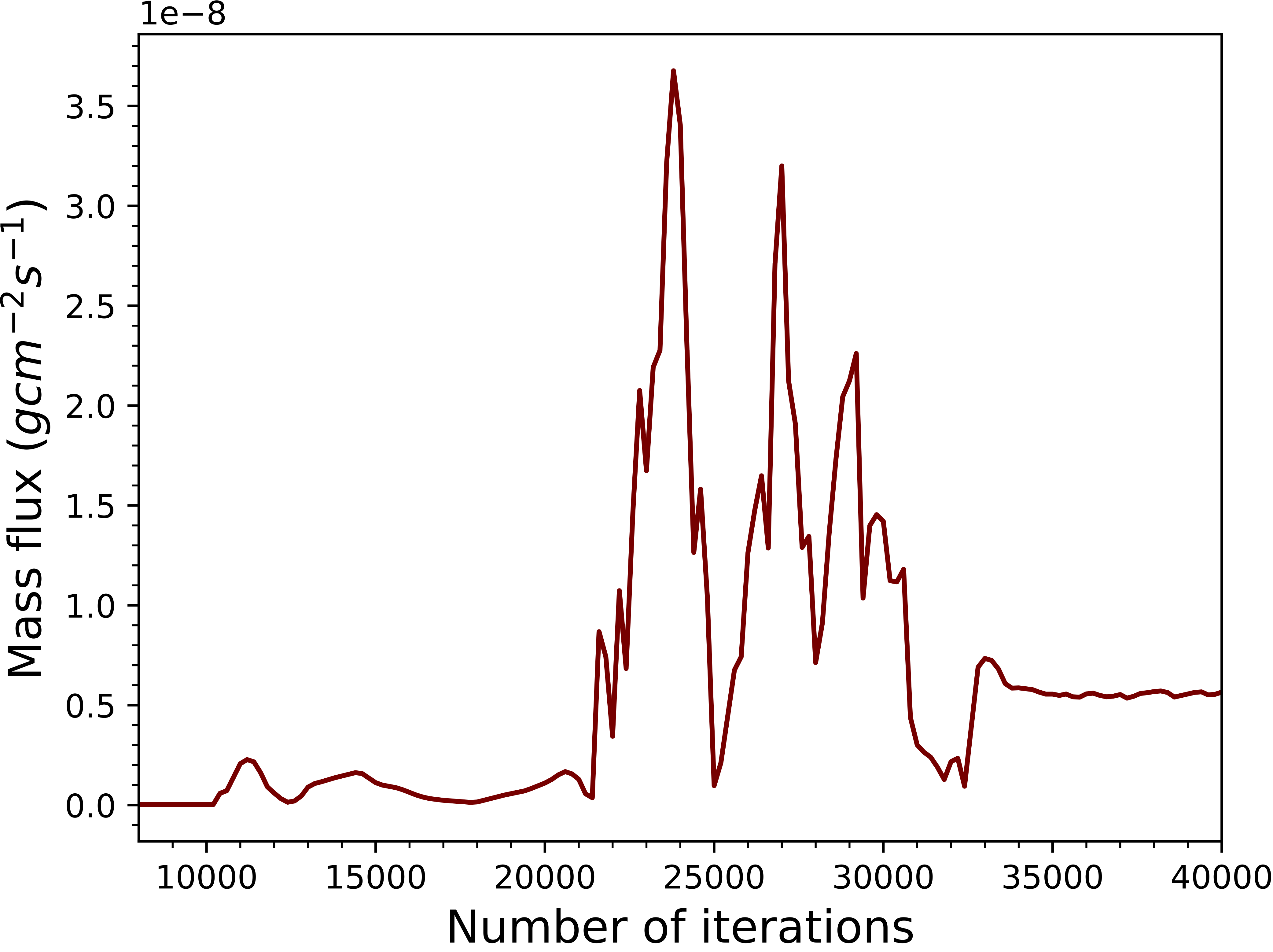}
  \caption{Mass flux leaving the planet (calculated at $\{x,y,z\} = \{ 7.0,-3.0,0\}R_p $) as a function of number of iterations for the case of planetary outflow only (top) and Case-IV (CME \& flare case, bottom).}
  \label{fig:mass_flux}
\end{figure}

A steady state in our simulation means two consecutive outputs have the same solution or the solutions vary within an error (quasi-steady state). To clarify this, figure~\ref{fig:mass_flux} shows the mass flux $(\rho v)$ leaving the planet (at $\{x,y,z\} = \{ 7.0,-3.0,0\}R_p $)  as a function of the number of iterations. The top panel shows the case where only the planetary outflow is simulated. We see that after about 4000 iterations, the mass flux is constant, indicating that steady state was achieved. The bottom panel of figure~\ref{fig:mass_flux} is similar to the upper panel, but now we consider Case-IV (CME \& flare case). Note that after 34000 iterations, there is a small variation in the mass flux, indicating that the simulation has reached a quasi-steady state. 
In this case, we calculated all the properties of the simulations (e.g., mass-loss rate) at the last iteration at 40000.

\bsp	
\label{lastpage}
\end{document}